\documentclass[twocolumn]{aastex63}
\usepackage[english]{babel}
\usepackage[utf8]{inputenc}
\usepackage{verbatim}
\usepackage{amsmath}
\usepackage{amsfonts}
\usepackage{amssymb}
\usepackage{graphicx}

\usepackage{empheq}
\usepackage{mathtools}
\usepackage{mathbbol}
\usepackage{fancyhdr}

\hypersetup{linkcolor=cyan,citecolor=cyan,filecolor=cyan,urlcolor=cyan}

\begin{document}

\title{\large Interplay between $\Delta$ Particles and Hyperons in Neutron Stars}

\author{Patricia Ribes}
\affiliation{Departament de F\'{\i}sica Qu\`antica i Astrof\'{\i}sica and Institut de Ci\`encies del Cosmos
(ICCUB), Facultat de F\'{\i}sica, Universitat de Barcelona, Mart\'{\i} i Franqu\`es 1, 08028
Barcelona, Spain}
\author[0000-0002-0882-1570]{Angels Ramos}
\affiliation{Departament de F\'{\i}sica Qu\`antica i Astrof\'{\i}sica and Institut de Ci\`encies del Cosmos
(ICCUB), Facultat de F\'{\i}sica, Universitat de Barcelona, Mart\'{\i} i Franqu\`es 1, 08028
Barcelona, Spain}
\author[0000-0003-2304-7496]{Laura Tolos}
\affiliation{Institut f\"ur Theoretische Physik, Goethe Universit\"at Frankfurt, Max von Laue Strasse 1, 60438 Frankfurt, Germany;\\
Frankfurt Institute for Advanced Studies,  Goethe Universit\"at Frankfurt, \\ Ruth-Moufang-Str. 1, 60438 Frankfurt am Main, Germany;   \\
Institute of Space Sciences (ICE, CSIC), Campus UAB, Carrer de Can Magrans, 08193, Barcelona, Spain; and \\
Institut d' Estudis Espacials de Catalunya (IEEC), 08034 Barcelona, Spain}
\author{Claudia Gonzalez-Boquera}
\affiliation{Departament de F\'{\i}sica Qu\`antica i Astrof\'{\i}sica and Institut de Ci\`encies del Cosmos
(ICCUB), Facultat de F\'{\i}sica, Universitat de Barcelona, Mart\'{\i} i Franqu\`es 1, 08028
Barcelona, Spain}
\author[0000-0002-4156-3067]{Mario Centelles}
\affiliation{Departament de F\'{\i}sica Qu\`antica i Astrof\'{\i}sica and Institut de Ci\`encies del Cosmos
(ICCUB), Facultat de F\'{\i}sica, Universitat de Barcelona, Mart\'{\i} i Franqu\`es 1, 08028
Barcelona, Spain}


\begin{abstract}
We analyze the effects of including $\Delta(1232)$ isobars in an equation of state (EoS) for cold, $\beta$-stable neutron star matter, employing relativistic nuclear mean field theory. The selected EoS reproduces the properties of nuclear matter and finite nuclei and, in the astrophysical context, allows for the presence of hyperons in neutron stars having masses larger than 2$M_{\odot}$. We find that the composition and structure of neutron stars is critically influenced by the addition of the $\Delta$ isobars, which allows us to constrain their interaction with the meson fields taking into account astrophysical information. Imposing that the EoS is stable and ensures the existence of 2$M_{\odot}$ neutron stars, as well as requiring agreement with data of $\Delta$ excitation in nuclei, we find that, in the absence of other mechanisms stiffening the EoS at high densities, the interaction of the $\Delta$ isobars with the sigma and omega meson fields must be at least 10\% stronger than that of the nucleons. Moreover, the neutron star moment of inertia turns out to be sensitive to the presence of $\Delta$ isobars, whereas
the inclusion of $\Delta$ isobars in the EoS allows for smaller stellar radii and for a lower value of the tidal deformability consistent with the analysis of the GW170817 merger event. 
\end{abstract}

  

\section{Introduction}
Neutron stars (NSs) are the most compact objects known without an event horizon \citep{Shapiro:1983du}. Their cores contain strongly interacting matter that is several times denser than the matter at the center of nuclei.
%
%
The properties of NSs primarily depend on the equation of state (EoS) of 
isospin-rich strongly interacting cold dense matter. Even though the EoS around saturation density can be determined from several experimental sources, mainly from the properties of nuclei and from the analysis of heavy-ion collisions, the large densities at the center of NSs have not been reached experimentally, and the characteristics of their innermost core remain largely unknown.

In order to model the EoS for predicting the structure and composition of matter in these extreme conditions, a variety of phenomenological theories, which take into account effective particle interactions, phase transitions and general relativity, have been developed, see e.g. \cite{Shapiro:1983du,glen,haensel}. 
By providing unique constraints on the tidal deformability of NSs, the recent GW170817 detection by the LIGO-Virgo collaboration of gravitational waves from the merger of a binary NS system \citep{TheLIGOScientific:2017qsa,Abbott:2018exr,Abbott:2018wiz} has stimulated further interest to explore the sensitivity of the NS properties on the EoS at large baryon density and isospin asymmetry. Since the tidal effects are strongly dependent on the stellar compactness, a measurement of the tidal deformability offers insights into the underlying EoS and the possibility to discriminate among EoSs that predict similar maximum NS masses but different NS radii.

The composition of a NS is driven by the so-called $\beta$-equilibrium condition, which establishes equilibrium among weak interaction processes, imposing charge neutrality and baryon number conservation. 
A NS consists mostly of neutrons, protons, and electrons. However, due to the large values of the nucleon chemical potential at the large inner densities, the conversion of nucleons to hyperons is energetically favorable. The appearance of hyperonic degrees of freedom relieves the Fermi pressure exerted by  baryons and makes the EoS softer, which generally leads to a reduction of the maximum possible mass  of NSs below $2M_{\odot}$, in disagreement with some observations \citep{2010Natur.467.1081D,2013Sci...340..448A}, especially with the very recent value of $2.17^{+0.11}_{-0.10}$ solar masses reported for the millisecond pulsar J0740+6620 in \cite{Cromartie:2019kug}.
Nevertheless, it is still possible to fine tune the models to fulfill the conditions imposed by recent observations and many theoretical studies regarding the presence of hyperons in the core of neutron stars have been conducted to achieve this goal, see \cite{Chatterjee:2015pua} and references therein. In this respect, relativistic mean field (RMF) models are especially powerful because they offer an effective and practical way to deal with the intricacies of the nuclear many-body problem at the high densities present in NSs \citep{glen,haensel}. These models rely on a few parameters---essentially meson-baryon coupling constants---that reproduce well-known properties of ordinary nuclei and assume plausible extrapolations to higher densities within a natural covariant framework. 

Comparatively, less work has been done to study the presence of $\Delta$(1232) isobars in NSs, as early studies indicated that these particles would only appear at much higher densities than the typical ones in NS cores \citep{giant}. A strong correlation between the appearance of $\Delta$ isobars and the $\rho$-meson coupling constant or, relatedly, the value of the symmetry energy in strongly interacting matter was pointed out. This type of study has been revisited in recent years \citep{Drago,Cai2015CriticalStars,Zhu:2016mtc,Li:2018qaw} in the light of the recent tight constraints obtained on the symmetry energy at saturation density and its slope, permitting a better extrapolation of the models to describe the dense and highly isospin-asymmetric conditions of NS matter. 
Indeed, the recent theoretical studies have shown that $\Delta$ particles may appear between 2 to 3 times nuclear matter saturation density, in competition with the onset of hyperons, and still have the potential to support $2M_{\odot}$ stars and to produce canonical $1.4M_{\odot}$ stars with relatively small radii below $\sim$13 km, as suggested by the latest astrophysical analyses, see e.g.~\cite{Fortin:2014mya,Ozel:2015fia,Ozel:2016oaf,Nattila:2015jra,Lattimer:2015nhk,Abbott:2018exr}. The presence of $\Delta$ baryons may also lead to reduced tidal deformabilities \citep{,Li:2019tjx} and bring them closer to the values measured from the gravitational waves emitted in the binary NS merger GW170817 \citep{TheLIGOScientific:2017qsa,Abbott:2018exr,Abbott:2018wiz}. The situation is, however, far from being settled. For instance, a recent study based on the quark-meson-coupling model, in which the $\Delta$-nucleon mass splitting emerges from a spin- and density-dependent one-gluon-exchange interaction, predicts a strongly repulsive $\Delta$ potential in dense matter, thus preventing the $\Delta$ isobars from appearing at all in NSs \citep{Motta:2019ywl}.

The aim of the present work is to analyze the interplay between $\Delta$ particles and hyperons in NSs and the implications for the properties of the star compared with the astrophysical observations, using the EoS of the recent fine-tuned RMF model FSU2H of \cite{angels,pasa}. This model allows for the presence of hyperons in NSs, while reproducing the properties of infinite nuclear matter and finite nuclei, the EoS constraints deduced from heavy-ion collisions, the symmetry energy and its density slope, and is compatible with the $2M_{\odot}$ observations as well as with relatively small radii for canonical NSs of $1.4M_{\odot}$ (also see \cite{Negreiros:2018cho} for an application of the model to cooling simulations). We will show that the presence of $\Delta$ isobars, which competes with that of hyperons, does not compromise the good features of the model for reasonable values of the $\Delta$ coupling parameters.
Indeed, we find that for compact stars that harbor strangeness in their dense interiors it
is important to take into account the $\Delta$ degrees of freedom in the stellar composition in order to satisfy at the same time the observational constraints of massive NSs, small stellar radii, and the tidal deformability values obtained by LIGO-Virgo from the GW170817 event. We also find that the consideration of these astrophysical bounds favors a $\sim$10\%--30\% deeper potential for the $\Delta$ isobars than for the nucleon in nuclear matter at normal density, an aspect of the $\Delta$ interaction that thus far is not fully resolved from the laboratory experiments \citep{Ericson:1988gk,Horikawa:1980cv,Nakamura:2009iq,Lehr:1999zr,OConnell:1990njm}. Finally, we have noticed that the moment of inertia of NSs is sensitive to the presence of the $\Delta$ baryons. The latter contribute to improve the agreement with recent EoS-independent constraints \citep{Landry:2018jyg} on the moment of inertia of the primary component of the double pulsar PSR J0737-3039.

The paper is organized as follows.  In Section \ref{sec:eos} we describe the main features of the model of the equation of state, whereas in Section \ref{sec:tov} we summarize the calculation of the mass, radius and tidal deformability of NSs. In Section \ref{sec:results} we discuss our results and our conclusions are given in Section \ref{sec:conclusions}.

\section{Equation of state}
\label{sec:eos}

The RMF theory \citep{Serot:1984ey} provides a Lorentz-covariant description of the microphysics of the EoS of dense matter \citep{glen}. In this work we study the properties of NSs by computing the EoS with the 
hyperonic RMF model FSU2H that we introduced in \cite{angels,pasa} 
supplemented by the $\Delta$(1232) degrees of freedom.
Altogether, the model contains the octet of the spin-$1/2$ light baryons ($n$, $p$, $\Lambda$, $\Sigma^{-,0,+}$, $\Xi^{-,0}$), 
as in the FSU2H case, plus the quartet of the spin-$3/2$ $\Delta$ baryons ($\Delta^{-,0,+,++}$). 
In the RMF theory the baryons are represented in terms of Dirac spinors and 
the interactions among the baryons are described through the exchange of virtual mesons.
The mesons of our model comprise a scalar-isoscalar $\sigma$ meson that provides the baryon-baryon attraction, a 
vector-isoscalar $\omega$ meson that provides the repulsion among baryons, a vector-isovector $\rho$ 
meson that accounts for the isospin dependence of the interaction, and a vector-isoscalar $\phi$ meson that couples to the hyperons. Including electrons and muons for the leptons present in cold catalyzed neutron star matter, the Lagrangian density can be written~as 
\begin{eqnarray} \label{eq:L}
\mathcal{L} &=& \sum_{b}\mathcal{L}_{b}  + \sum_{\Delta}\mathcal{L}_{\Delta} + \sum_{l} \mathcal{L}_l + \mathcal{L}_m ,
\end{eqnarray}
where (we use units $\hbar=c=1$ throughout this work):
\begin{eqnarray} \label{eq:lags}
\mathcal{L}_b &=& \bar{\Psi}_b(i\gamma_{\mu}\partial^{\mu} - m_b + g_{\sigma b} \sigma - g_{\omega b}\gamma_{\mu}\omega^{\mu}   \nonumber \\ 
&& \mbox{} - g_{\rho b}\gamma_{\mu}\vec{I}_{b}\!\cdot\!\vec{\rho}^{\,\mu}- g_{\phi b}\gamma_{\mu}\phi^{\mu})\Psi_b , \nonumber \\[2mm]
\mathcal{L}_{\Delta} &=& \bar{\Psi}_{\Delta} (i\gamma_{\mu}\partial ^{\mu} - m_{\Delta} + g_{\sigma \Delta}\sigma -g_{\omega \Delta} \gamma_{\mu}\omega^{\mu} \nonumber \\
&& \mbox{} - g_{\rho \Delta}\gamma_{\mu}\vec{I}_{\Delta}\!\cdot\!\vec{\rho}^{\,\mu})\Psi_{\Delta} , \nonumber \\[2mm]
\mathcal{L}_l &=& \bar{\psi}_l (i\gamma_{\mu}\partial^{\mu} - m_l )\psi_l , \nonumber \\[2mm]
\mathcal{L}_m &=& \frac{1}{2}\partial_{\mu}\sigma\partial^{\mu}\sigma - \frac{1}{2}m_{\sigma} ^2 \sigma^2 - \frac{\kappa}{3!}(g_{\sigma N}\sigma)^3 - \frac{\lambda}{4!}(g_{\sigma N}\sigma)^4 \nonumber \\ 
&& \mbox{} - \frac{1}{4}\Omega ^{\mu \nu} \Omega_{\mu \nu} + \frac{1}{2}m_{\omega} ^2\omega_{\mu}\omega^{\mu} 
   + \frac{\zeta}{4!} (g_{\omega N }\omega_{\mu}\omega^{\mu})^4 \nonumber \\
&& \mbox{} - \frac{1}{4} \vec{R}^{\mu \nu}\!\cdot\!\vec{R}_{\mu \nu} + \frac{1}{2}m_{\rho}^2\vec{\rho}_{\mu}\!\cdot\!\vec{\rho}^{\,\mu} \nonumber \\
&& \mbox{} + \Lambda_{\omega} g_{\rho N}^2 \vec{\rho}_{\mu}\!\cdot\!\vec{\rho}^{\,\mu}g_{\omega N}^2 \omega_{\mu}\omega^{\mu} \nonumber \\ 
&& \mbox{} -\frac{1}{4}P^{\mu \nu}P_{\mu \nu} + \frac{1}{2} m_{\phi}^2\phi_{\mu}\phi^{\mu} .
\end{eqnarray} 
The subscripts $b$ and $\Delta$ run over the baryon octet and the $\Delta$ quartet, respectively, 
whereas the subscript $l$ runs over the electrons and muons. 
The field $\Psi_b$ is the baryon Dirac field, $\Psi_{\Delta} $ is the Rarita-Schwinger field for the $\Delta$ isobars,
and $\psi_l$ is the lepton Dirac field. 
The strong interaction couplings between the different baryons and mesons are denoted by $g$, whereas
the $\vec{I}$ vector stands for the isospin operator.
The $\mathcal{L}_m$ term contains the contribution from the free meson fields.
The quantities $\Omega_{\mu \nu} $, $\vec{R}_{\mu \nu}$ and 
$P_{\mu\nu}$ denote the strength tensors of the $\omega$, $\rho$ and $\phi$ meson
fields, respectively. The $\kappa$ and $\lambda$ couplings associated with
the non-linear self-interactions of the $\sigma$ field \citep{Boguta:1977xi} are responsible for reducing
the value of the nuclear matter incompressibility to within the experimental range.
The main effect of the $\zeta$ self-coupling of the $\omega$ meson is to soften the EoS in the high-density sector \citep{Mueller:1996pm}.
Finally, the $\Lambda_{\omega}$ coupling of the quartic isoscalar-isovector interaction between the $\omega$ and $\rho$ mesons
allows one to adjust the density dependence of the nuclear symmetry energy~\citep{Horowitz:2000xj,Chen:2014sca,Chen:2014mza}. 

From the above Lagrangian one derives in the standard way the energy density of the system.
Working in the mean-field approximation, where the meson fields are replaced
by their ground-state expectation values, namely, $ \bar{\sigma}= \langle \sigma \rangle$, $ \bar{\omega}= \langle \omega^0 \rangle$, 
$ \bar{\rho}= \langle \rho^0_3 \rangle$, and $ \bar{\phi}= \langle \phi^0 \rangle$ (under time-reversal symmetry
and charge conservation, only the time components of the vector fields and the third component in isospin space of the $\vec{\rho}$ field contribute),
the result is \citep{angels,pasa}  
\begin{eqnarray}
\epsilon &=& \sum_b \epsilon_b + \sum_{\Delta} \epsilon_{\Delta} + \sum_l \epsilon_l + \frac{1}{2}m_{\sigma}^2\bar{\sigma}^2 + \frac{1}{2}m_{\omega}^2\bar{\omega}^2  \nonumber \\ 
&&\mbox{} +\frac{1}{2}m_{\rho}^2 \bar{\rho}^2 + \frac{1}{2}m_{\phi}^2 \bar{\phi}^2 + \frac{\kappa}{3!} (g_{\sigma N}\bar{\sigma})^3 + \frac{\lambda}{4!}(g_{\sigma N}\bar{\sigma})^4  \nonumber \\
&&\mbox{} +\frac{\zeta}{8} (g_{\omega N}\bar{\omega})^4 + 3\Lambda_{\omega}(g_{\rho N}g_{\omega N}\bar{\rho}\,\bar{\omega})^2 .
\end{eqnarray}
The contributions $\sum_b \epsilon_b$, $\sum_{\Delta} \epsilon_{\Delta} $, and $\sum_l \epsilon_l $ from
the baryons of the octet, the $\Delta$ isobars, and the leptons are obtained as analytical expressions 
of the Fermi momenta of the particles and their effective masses~($ m_{b}^* = m_{b }- g_{\sigma b}\bar{\sigma}$,
$ m_{\Delta}^* = m_{\Delta }- g_{\sigma \Delta}\bar{\sigma}$, $m_l^*=m_l$), cf.\ \cite{angels,pasa}. 

The pressure of the system can be obtained as
\begin{equation}
P = \sum_i \mu_i n_i - \epsilon ,
\end{equation} 
where $n_i$ are the densities of the different particles and $\mu_i$ are their chemical potentials. The latter take the expressions 
\begin{eqnarray}
\mu_b & =& (k_{Fb}^2+{m^*_b}^2)^{1/2} + g_{\omega b }\bar{\omega} + g_{\rho b}I_{3b}\bar{\rho} + g_{\phi b } \bar{\phi}, \\  \label{eq:chemdelta}
\mu_{\Delta} & =& (k_{F\Delta}^2+{m^*_\Delta}^2)^{1/2} + g_{\omega \Delta }\bar{\omega} + g_{\rho \Delta}I_{3\Delta}\bar{\rho},\\
\mu_l & =& (k_{Fl}^2+m_l^2)^{1/2} ,
\end{eqnarray}
where $k_{Fb}$, $k_{F\Delta}$ and $k_{Fl}$ are the Fermi momenta of
the baryons, $\Delta$ isobars and leptons, respectively, and $I_{3}$ denotes the third component of the isospin operator, 
with the convention that for protons we have $I_{3p}= +1/2$.
The conditions of charge neutrality and $\beta$-equilibrium of the matter in the NS core impose the following 
requirements on the particle densities $n_i$ and chemical potentials~$\mu_i$:
\begin{eqnarray}
0 & = & \sum_{i=b,\Delta,l} q_i n_i, \label{eq:balance}\\
\mu_i & = & b_i \mu_n - q_i \mu_e, \label{eq:mu}
\end{eqnarray}
where $q_i$ and $b_i$ are the electrical charge and baryon number of particle $i$. For a given baryon density $n$ such that  
\begin{equation}
n = \sum_{i=b,\Delta} n_i , \label{eq:totaln}
\end{equation}
the equations of motion for the meson fields and the different particle species are solved self-consistently, under the 
restrictions presented in Eqs.\ (\ref{eq:balance})--(\ref{eq:totaln}), in order to obtain the chemical potential and the 
corresponding density of each of the species. One is thereby able to determine the composition of the stellar matter as a function of 
the baryonic density and the relation between the energy density and pressure (EoS) of the system. 


For the nucleon and hyperon interactions we adopt the RMF parametrization FSU2H \citep{angels,pasa}, with 
the values of the coupling constants and masses given in Table 1 of \cite{pasa}.
This model predicts realistic properties for ordinary nuclear matter, with a saturation density 
$n_0=0.1505$ fm$ ^{-3}$, saturation energy per particle $E/A=-16.28$ MeV, incompressibility 
$K=238$ MeV, symmetry energy $E_\textrm{sym}= 30.5$ MeV, and density slope
of the symmetry energy $L= 3n_0 (\partial E_\textrm{sym}(n)/\partial n)\vert_{n_0}= 44.5$ MeV. It
also provides a good description of the ground-state properties of atomic nuclei \citep{angels,pasa}.
In the hyperonic sector, we have determined \citep{angels,pasa} the values of the couplings between the 
hyperons and the vector mesons using SU(3)-flavor symmetry, the vector dominance model and ideal mixing for
the $\omega$ and $\phi$ mesons \citep{Schaffner:1995th,Banik:2014qja,Miyatsu:2013hea,Weissenborn:2011ut,Colucci:2013pya}.
In the case of the coupling of the $\Lambda$-hyperon to the $\phi$-meson, we have allowed for a reduction 
of 20\% from its SU(3) value in order to reproduce the $\Lambda\Lambda$ bond energies \citep{Ahn:2013poa}.
Regarding the couplings of the hyperons to the scalar $\sigma$-meson, we have adjusted them for describing
laboratory data on hypernuclei, in particular the hyperon optical potential. 
The values that we consider in the present work for the couplings $g_{\sigma \Delta}$, $g_{\omega \Delta}$, 
and $g_{\rho \Delta}$ between the $\Delta$ resonance and the mesons are discussed later in the section of results. 
As a final condition on the EoS with the $\Delta$ degrees of freedom, 
we will demand the resulting EoS to be stable (i.e., $d P / d \epsilon > 0 $ for all considered densities)
and stiff enough as to ensure the existence of NSs with masses larger than 2$M_{\odot}$,
in agreement with recent observations. These two restrictions will have a crucial influence on limiting the possible
values for the $\Delta$ couplings.

The calculation of the structure and properties of a NS, such as the mass and the radius,
requires the knowledge of the relation between pressure and energy density not only in the uniform matter 
of the liquid core but also in the solid crust of the star. The EoS for the inner crust 
with the FSU2H interaction has been recently computed in \cite{Providencia:2018ywl}.
This allows us to describe with FSU2H the EoS for NSs in a unified way at the level 
of the core and the inner crust of the star. For the external layers of the crust, i.e., the outer crust, we employ the widely used 
EoS of Baym-Pethick-Sutherland (BPS) \citep{Baym:1971pw}, which is strongly constrained by nuclear physics data.


\section{Stellar structure}
\label{sec:tov}


The structure of NSs is affected by general relativistic effects. Imposing hydrodynamic equilibrium of the star within 
General Relativity yields the set of differential equations known as Tolman-Oppenheimer-Volkoff (TOV) equations:
\begin{eqnarray}
\frac{dP(r)}{dr} &=& -\frac{G}{r^2}\left[\epsilon(r) + P(r)\right] \left[m(r) + 4\pi r^3 P(r) \right] \nonumber \\
& & \times \left[ 1- \frac{2Gm(r)}{r}\right]^{-1}\\
\frac{dm(r)}{dr} &=& 4\pi r^2 \epsilon (r),
\end{eqnarray}
where $G$ is the gravitational constant, $r$ is the distance to the center of the star and $m(r)$ is the mass enclosed within a 
sphere of radius $r$. Once the EoS of stellar matter is supplied, the TOV equations can be 
integrated from the origin with initial conditions $m(0) = 0$ and an arbitrary 
value for the central energy density $\epsilon(0) = \epsilon_c$ until the pressure vanishes. The point $R$ where $P(R)=0$ defines the 
radius and total mass, $M=m(R)$, of the star. 

Pressure plays a fundamental role in the determination of the structure of relativistic stars,
as it is ultimately responsible for the existence of a limiting mass in these objects \citep{glen}. 
The specific values of the mass and radius of NSs are linked to the details 
of the EoS, a fact that offers a way to test the validity of the different models 
(and to constrain their parameters), by requiring that they can reproduce the observed heavy NS masses 
of around 2$M_{\odot}$ \citep{2010Natur.467.1081D,2013Sci...340..448A} and predict radii below
$\sim$13~km for canonical NSs of $1.4 M_\odot$, as suggested by recent astrophysical
extractions of NS radii \citep{Fortin:2014mya,Ozel:2015fia,Nattila:2015jra,Ozel:2016oaf,Lattimer:2015nhk,Abbott:2018exr}.
In particular, the maximum NS mass predicted by an EoS strongly depends on the stiffness of the core, 
which is known to become softer with the appearance of hyperons \citep{angels} and $\Delta$ isobars \citep{Drago}.


In a coalescing binary system of two NSs, the component stars experience a quadrupole deformation due to the gravitational tidal field
caused by their compact companion. The tidal deformation effect on a NS is characterized by the so-called tidal deformability $\lambda$,
which is defined as the ratio between the induced quadrupole moment and the external tidal field \citep{Flanagan:2007ix}. The tidal deformability
$\lambda$ can be expressed in terms of the dimensionless second tidal Love number $k_2$ and 
the radius of the star as follows \citep{Flanagan:2007ix,Hinderer:2007mb}:
\begin{equation}\label{lamb}
 \lambda = \frac{2}{3} \frac{R^5}{G} k_2 .
\end{equation}
The Love number $k_2$ is calculated by solving an additional differential equation
self-consistently with the integration of the TOV equations \citep{Hinderer:2007mb,Hinderer:2009ca,Postnikov:2010yn}. 
The global tidal effect of the two NSs in an inspiraling binary is described by the mass-weighted tidal deformability 
\citep{Flanagan:2007ix,Hinderer:2009ca,Damour:2009vw}:
\begin{equation}\label{tlamb}
\tilde{\lambda} = \frac{1}{26} \left( \frac{M_1+12 M_2}{M_1} \lambda_1 + \frac{M_2+12M_1}{M_2}\lambda_2\right) ,
\end{equation}
where $\lambda_1$ and $\lambda_2$ are the tidal deformabilities of the two component stars and $M_1$ and $M_2$ are their masses (if $M_1=M_2$, then $\tilde{\lambda}= \lambda_1 = \lambda_2$). The same quantity can be expressed in dimensionless form as
\begin{equation}\label{TLambda}
 \tilde{\Lambda} = \frac{32}{G^4 (M_1+M_2)^5} \, \tilde{\lambda} .
\end{equation}
During the early stages of an inspiral, the phase of the gravitational wave signal is determined to leading
order by $\tilde{\Lambda}$ \citep{Flanagan:2007ix,Favata:2013rwa}. As the tidal deformabilities depend on the
EoS---recall Eq.~(\ref{lamb}), measurements of $\tilde{\Lambda}$ from the gravitational waves emitted
in a binary NS merger, such as the GW170817 event \citep{TheLIGOScientific:2017qsa,Abbott:2018wiz}, are a potential
probe of the underlying EoS of the matter of compact stars.

\section{Results and discussion}
\label{sec:results}

The EoS presented in Sec.~\ref{sec:eos} depends on the strong interaction couplings  of the baryons to the meson fields (denoted by $g_{m b}$ with $m=\sigma,\omega,\rho,\phi$) and on the parameters of the self-interactions of these  fields ($\kappa$, $\lambda$, $ \zeta$ and $\Lambda_{\omega}$).  We employ the values of the FSU2H model \citep{angels,pasa} (see Table 1 of \cite{pasa}), with the proviso that the incorporation of the $\Delta$ degrees of freedom introduces three new free parameters, i.e., $g_{\sigma \Delta}$, $g_{\omega \Delta}$ and $g_{\rho\Delta}$.  As is customary, in the discussion that follows we will treat the values of these parameters in terms of the ratios to the meson-nucleon couplings:
\begin{align}
x_{\sigma \Delta} = \frac{g_{\sigma \Delta}}{g_{\sigma N}} \qquad   x_{\omega \Delta} = \frac{g_{\omega \Delta}}{g_{\omega N}}  \qquad x_{\rho \Delta} = \frac{g_{\rho \Delta}}{g_{\rho N}}.
\end{align}

Even though the couplings of the $\Delta$ isobars with the meson fields are poorly constrained due to the limited existence of experimental observations, some information is available that may be used to restrict their values within some range. While most of the phenomenological analyses advocate for an attractive $\Delta$-nucleus potential, no consensus has been reached on its actual size. It has been claimed to be only a few tenths of MeV at the nuclear surface from pion-nucleus scattering reactions \citep{Ericson:1988gk,Horikawa:1980cv,Nakamura:2009iq}, or practically the same as that of the nucleon from photo-absorption reactions \citep{Lehr:1999zr}, or up to a 30\% larger from inclusive electron-nucleus scattering data \citep{OConnell:1990njm}.  Moreover, as mentioned e.g.\ in \cite{Drago}, phenomenological analyses of electron-nucleus reactions in the region of the $\Delta$ excitation within a relativistic quantum hadrodynamics mean field model provided the following constraint \citep{WEHRBERGER1989797}:
\begin{equation} \label{eq:experiment}
0 \lesssim x_{\sigma \Delta} - x_{\omega \Delta} \lesssim 0.2 \, .
\end{equation}
Taking this phenomenology into consideration, we will explore how the composition and structure of NSs is affected by small modifications of the coupling of the $\Delta$ baryon to the $\sigma$ and $\omega$ mesons around values similar to those of the nucleon, i.e., taking $x_{\sigma\Delta}$ and $x_{\omega\Delta}$ within the range $[0.8-1.2]$. No information is available for the coupling of the $\rho$ meson to the $\Delta$ and, for the time being, we will adopt the value $x_{\rho\Delta}=1$, although the star properties for large variations of this unconstrained parameter will also be explored.

\begin{figure}[t]
\centering
\includegraphics[width=0.45\textwidth]{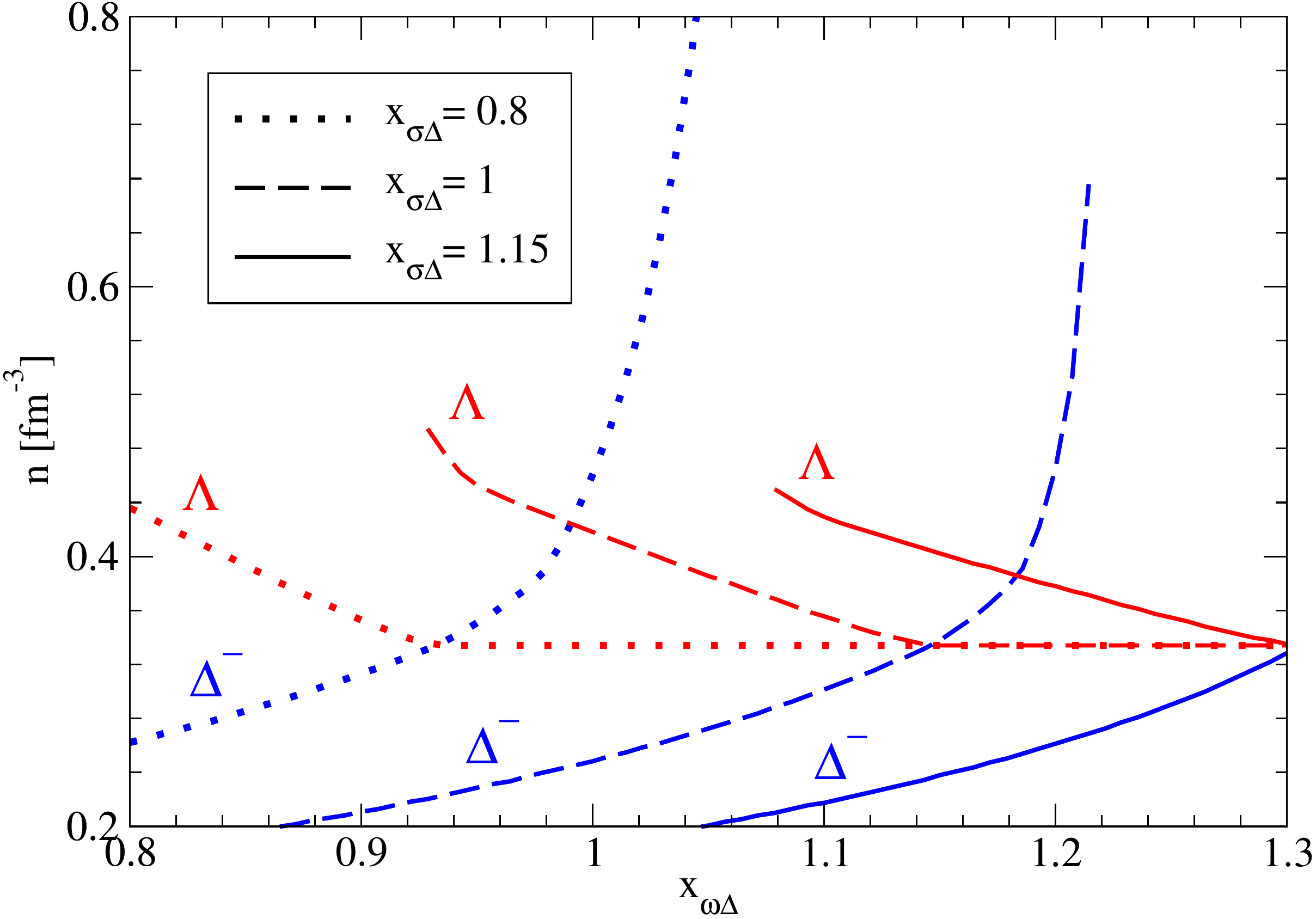}
\caption{Threshold densities of the $\Lambda$ hyperon and the $\Delta^-$ particle as function of $x_{\omega\Delta}$ for three different values of $x_{\sigma \Delta}$, fixing $x_{\rho \Delta} = 1$.
}
\label{fig:onsets}
\end{figure}

 In Fig.~\ref{fig:onsets}, we study the dependence of the onset density (the lowest density at which a particle first appears) of the $\Delta^-$ and the $\Lambda$ hyperon in $\beta$-stable matter as a function of $x_{\omega \Delta}$, for different values of $x_{\sigma \Delta}$ and taking  $x_{\rho \Delta} = 1$. Among the four $\Delta$ particles, the $\Delta^-$ is the first one to appear since, being negatively charged, it can replace a neutron and an electron at the top of their Fermi seas. The $\Lambda$ is the hyperon that first appears, owing to its lowest mass and to the fact that the $\Sigma^-$ feels a repulsive interaction \citep{Kohno:2006iq}.

For all chosen values of $x_{\sigma \Delta}$, namely $x_{\sigma \Delta}=0.8$ (dotted lines), $x_{\sigma \Delta}=1$ (dashed lines), and $x_{\sigma \Delta}=1.15$ (solid lines), we find that the $\Delta^-$ appears at smaller densities than the $\Lambda$ hyperon for the smaller values of~$x_{\omega \Delta}$.  As the value of $x_{\omega\Delta}$ increases, the onset density of the $\Delta^-$ also increases and that of the $\Lambda$ decreases, leading to the existence of a certain $x_{\omega\Delta}$, for the given $x_{\sigma \Delta}$, at which both the $\Delta^-$ and the $\Lambda$ appear at the same density. For larger $x_{\omega\Delta}$, the $\Lambda$ hyperon appears before the $\Delta^-$, at 0.336 fm$^{-3}$, which corresponds to the onset density of the $\Lambda$ in the FSU2H model without $\Delta$ isobars  \citep{angels}. 
 We note that, if the constraint of Eq.~(\ref{eq:experiment}) is imposed, then we limit the range of validity of the curves in Fig.~\ref{fig:onsets} to cases in which the $\Delta^-$ always appears first, delaying the appearance of the $\Lambda$ hyperon at a higher density than in the absence of $\Delta$ isobars, in agreement with the outcome of \cite{Drago}. Note also that the curves are not very much extended to the left, i.e., $x_{\sigma\Delta} - x_{\omega\Delta}$ may not be reaching its upper boundary of $~$0.2, because we encounter situations in which $dP/d\epsilon < 0$ and the EoS becomes unstable.


The results of  Fig.~\ref{fig:onsets} can be easily understood from the dependence of the $\Delta$ baryon potential, defined as 
\begin{equation}
    U_\Delta = - g_{\sigma \Delta}\bar{\sigma} + g_{\omega \Delta }\bar{\omega} + g_{\rho \Delta}I_{3\Delta}\bar{\rho} ,
    \label{eq:pot_delta}
\end{equation}
on the $\sigma$ and $\omega$ fields [see also Eq.~(\ref{eq:chemdelta})]. If $g_{\sigma \Delta}$ increases, the potential of the $\Delta^-$ becomes
smaller, as the contribution of the $\sigma$ field is attractive, thus favoring the appearance of the $\Delta^- $ at smaller densities, while
large values of $g_{\omega \Delta}$ increase the $\Delta^- $ potential because of the repulsive $\omega$ field, contributing to the opposite effect.

\begin{figure}[t]
\centering 
\includegraphics[width=0.6\textwidth]{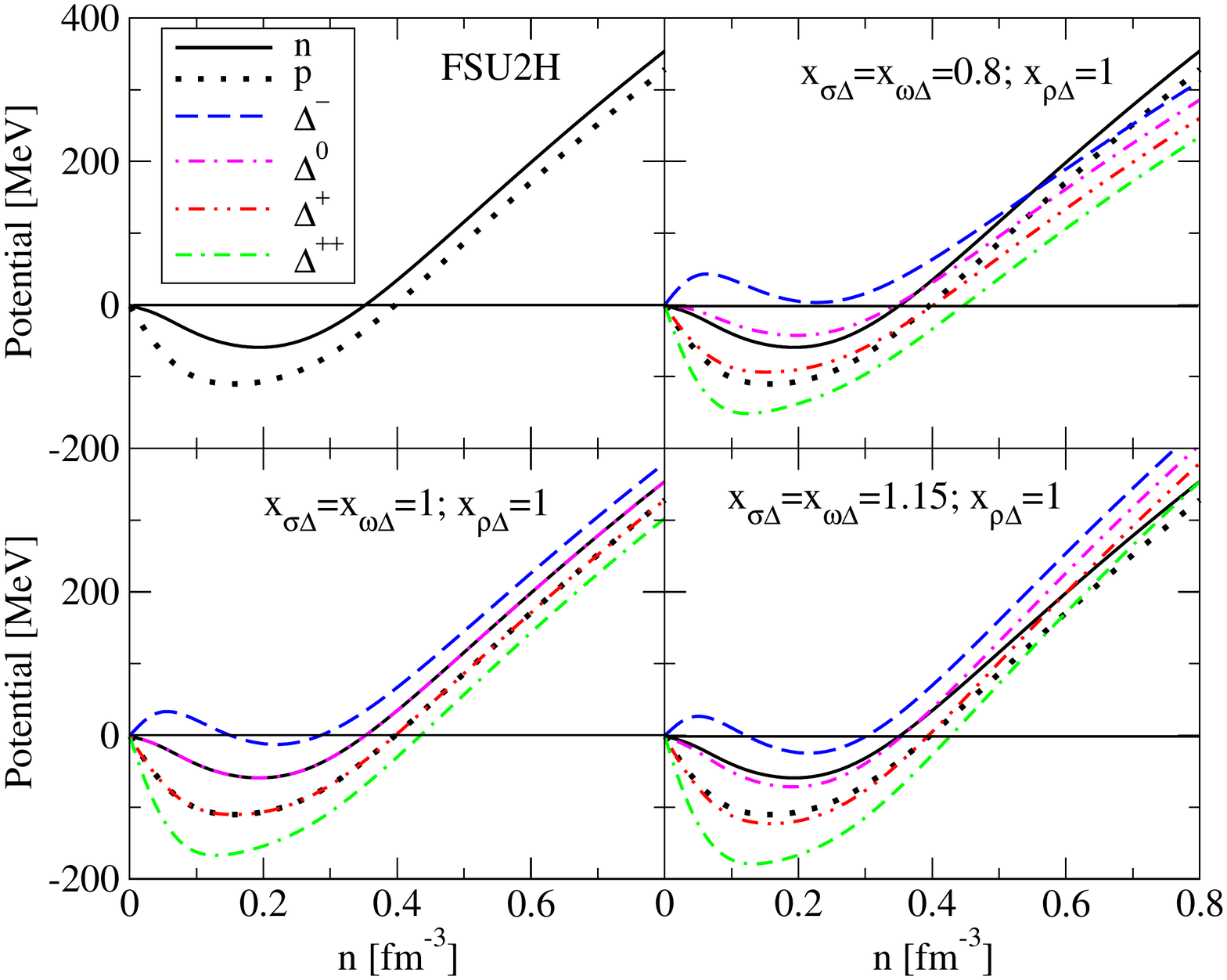}
\caption{Single-particle potentials of nucleons in pure neutron matter (upper left panel), compared with those of the $\Delta$ isobars, taking $x_{\sigma \Delta}=x_{\omega \Delta}=0.8$ (upper right panel), $x_{\sigma \Delta}=x_{\omega \Delta}=1$ (lower left panel), and $x_{\sigma \Delta}=x_{\omega \Delta}=1.15$ (lower right panel). We have taken the value $x_{\rho \Delta}=1$ in all cases.}
\label{fig:pot_neu}
\end{figure}

This is more clearly seen in Fig.~\ref{fig:pot_neu}, where the potentials of the nucleons ($U_N = - g_{\sigma N}\bar{\sigma} + g_{\omega N }\bar{\omega} + g_{\rho N}I_{3N}\bar{\rho}$) in pure neutron matter, shown in the upper left panel, are compared to those of the 
$\Delta$ particles ($\Delta^{-}$, $\Delta^{0}$, $\Delta^{+}$, and $\Delta^{++}$) in the other panels. The conditions of pure neutron matter resemble those found in $\beta$-stable NS matter.
Results are shown for three combinations of the ratio parameters:
$x_{\sigma \Delta}=x_{\omega \Delta}=0.8$ (upper right panel),
$x_{\sigma \Delta}=x_{\omega \Delta}=1$ (lower left panel),
and $x_{\sigma \Delta}=x_{\omega \Delta}=1.15$ (lower right panel). We have taken $x_{\rho \Delta}=1$ in all cases. 
As clearly seen in the upper left panel of Fig.~\ref{fig:pot_neu}, the nucleon potentials are negative for densities $n\lesssim 0.4$ fm$^{-3}$, due to the dominance of the attractive $\sigma$ field
in this region, and, conversely, they are positive at higher densities when the repulsive $\omega$ field overcomes the attractive $\sigma$ field.
Note that the potentials of the different charged baryons in each isospin family
differ due to the isovector term $g_{\rho}\bar{\rho}$ associated to the $\rho$ meson, which is negative in pure neutron matter.
Therefore, the proton potential is smaller than that of the neutron.
Similarly, when $x_{\sigma \Delta}=x_{\omega \Delta}=x_{\rho \Delta}=1$ (lower left panel of Fig.~\ref{fig:pot_neu}),
out of the four charged $\Delta$ states, the $\Delta^0$ and $\Delta^+$ potentials coincide with those of the neutron and the proton, respectively,
while the $\Delta^-$ state is the one that feels a higher potential and the $\Delta^{++}$ feels the lowest one. 
From Fig.~\ref{fig:pot_neu} we observe that if $x_{\sigma \Delta}=x_{\omega \Delta}=0.8$ (upper right panel of Fig.~\ref{fig:pot_neu}), the $\Delta^0$ and $\Delta^+$ potentials are in size more
moderate than the corresponding neutron and proton ones, being less attractive up to densities $n\sim 0.4$ fm$^{-3}$ and
less repulsive beyond that point. In particular, the smaller repulsion at higher densities is the
origin for the instabilities that are found in $\beta$-stable NS matter, as we will see. These instabilities are
delayed in density if the $\Delta$ couplings are the same as the nucleon ones but not enough to sustain NSs with 2$M_{\odot}$.
These massive NSs can be obtained when we take $x_{\sigma \Delta}=x_{\omega \Delta}=1.15$ (lower right panel of Fig.~\ref{fig:pot_neu}), which produces
more attractive $\Delta$ potentials at lower densities and more repulsive ones at higher densities.

\begin{figure}[t]
\centering 
\includegraphics[width=0.5\textwidth]{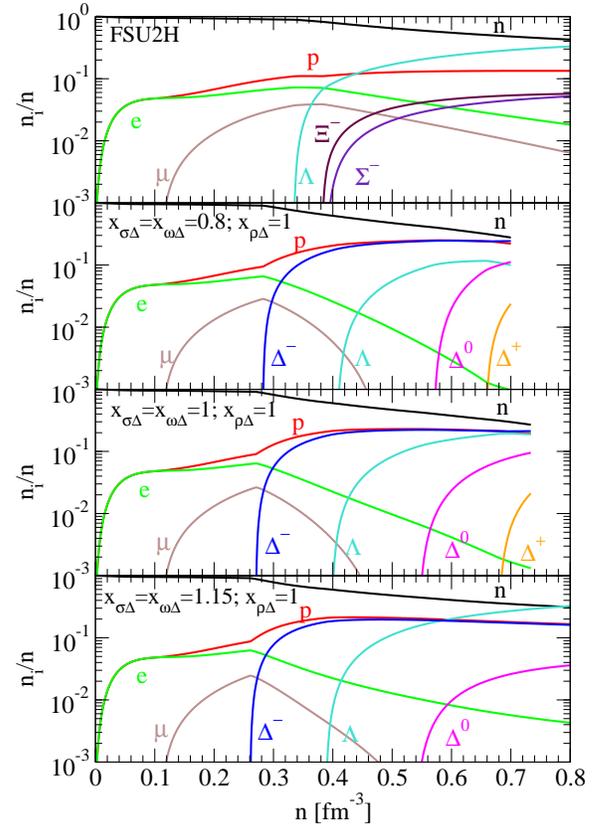}
\caption{Particle fractions as a function of the baryonic density (in fm$^{-3}$) within the FSU2H model including  hyperons. The upper panel shows the results for the model without $\Delta$ isobars. The remaining three panels correspond to the models with $\Delta$-meson couplings $x_{\sigma \Delta}=x_{\omega \Delta}=0.8$, $1$ and $1.15$, respectively, taking $x_{\rho \Delta}=1$.}
\label{fig:comp}
\end{figure}

In Fig.~\ref{fig:comp}  we compare the particle fractions in $\beta$-stable matter as functions of the baryon density for the FSU2H EoS without including $\Delta$ isobar degrees of freedom (first panel) and for $x_{\sigma \Delta}=x_{\omega \Delta}=0.8$ (second panel), 1 (third panel) and 1.15 (fourth panel), keeping in the latter three cases $x_{\rho \Delta} = 1$. One clearly sees the delay in the appearance of the $\Lambda$ hyperon caused by the presence of the $\Delta^-$. 
As the couplings $x_{\sigma \Delta}=x_{\omega \Delta}$ increase, the $\Delta^-$ and $\Delta^0$ occur at a lower density since their corresponding potential is lower due to the fact that, in this density region, the size of the attractive $\sigma$ field is still larger than that of the repulsive $\omega$ field. Conversely, the appearance of the $\Delta^+$ and $\Delta^{++}$ takes place at larger density as $x_{\sigma \Delta}=x_{\omega \Delta}$ is increased, since these two $\Delta$ components enter at densities for which the repulsive $\omega$ field dominates over the attractive $\sigma$ field.

In general, our results are consistent with the findings discussed in previous works, although with some noticeable differences.  In the case, for instance, of $x_{\sigma \Delta}=x_{\omega \Delta}=x_{\rho \Delta}=1$ we find an onset density for the $\Delta^-$ of 0.27~fm$^{-3}$, somewhat larger than the value 0.22~fm$^{-3}$ reported in \cite{Li:2018qaw} and sensibly smaller than the value 0.4~fm$^{-3}$ quoted in \cite{Drago}. The origin of these differences must be analyzed from the condition $\mu_n + \mu_e = \mu_{\Delta^-}$, which, at the onset density, can be written as $g_{\rho \Delta} \bar{\rho} + \sqrt{k_{Fn}^2+m_n^{*^2}}  + \mu_e =m_{\Delta^-}^*$. As discussed in \cite{Drago}, the fulfillment of this condition depends on a delicate balance between the value of the nuclear parameter $L$ (density slope of the symmetry energy), which controls the rise of the difference $\mu_n-\mu_p$ and, hence, of $\mu_e$, and the value of the term $g_{\rho \Delta} \bar{\rho}$, which is negative. A large $L$ value favors a lower onset density of the $\Delta^-$ as $\mu_e$ increases, but the negative field $\bar{\rho}$ will also be larger, with the opposing effect. One must also consider the particular value of the coupling strength $g_{\rho N}$. In general, models that have a larger $L$ value (implying a stiffer nuclear symmetry energy), also have a more moderate $g_{\rho N}$ value, see \cite{angels}, so that they may end producing $\Delta^-$ particles at relatively low densities, as is the case in \cite{Li:2018qaw}. Models with lower $L$ values (softer nuclear symmetry energy), which usually come with larger $g_{\rho N}$ strengths, give rise to larger $\Delta^-$ onset densities, as found here and in \cite{Drago}. It is thus fair to say that the model dependence situates the  appearance of the $\Delta^-$ in a density range $[0.2-0.4]$~fm$^{-3}$, also in line to what is found in the hyperon-free model of \cite{Cai2015CriticalStars}.
It is also interesting to notice that the $\Delta ^-$ partially replaces the role of the  $\Sigma^-$ and $\Xi ^-$ in compensating the positive charge of the protons, which, in the densities explored, leads to the absence of these hyperons in the composition of the star when the $\Delta$ isobars are considered. The shifting of the hyperon onset to higher densities when $\Delta$ isobars are present is a well-known fact \citep{Drago2014CanExist}, but there is also a clear model dependence in this delay, as the models discussed in  \cite{Drago,Li:2018qaw} permit the appearance of the $\Xi^-$ below $n=0.8$~fm$^{-3}$.

\begin{figure}
\centering 
\includegraphics[width=0.5\textwidth]{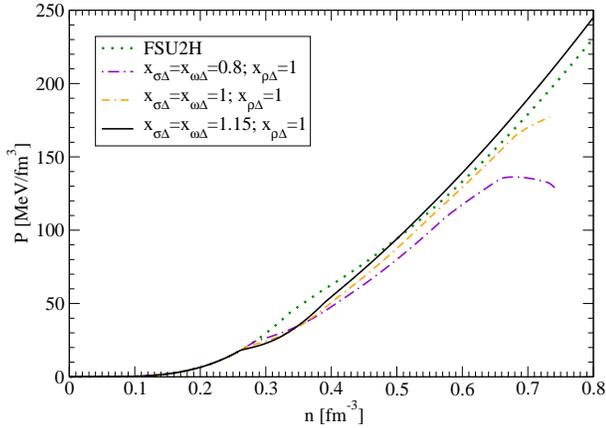}
\caption{Pressure versus baryon density of $\beta$-stable NS matter for the FSU2H model and various strengths of the $\Delta$-meson couplings.}
\label{fig:eos}
\end{figure}

In Fig.~\ref{fig:eos} we show the dependence of the pressure versus the baryon density in the case of $\beta$-stable NS matter for the four models discussed before. With respect to the model without $\Delta$ isobars, we clearly see that the case $x_{\sigma \Delta}=x_{\omega \Delta}=0.8$ becomes unstable beyond four times nuclear matter density and therefore is unable to sustain NS masses of 2$M_\odot$, as we will see. The instability is somewhat delayed for the case $x_{\sigma \Delta}=x_{\omega \Delta}=1$, whereas the case $x_{\sigma \Delta}=x_{\omega \Delta}=1.15$ is well behaved for all densities, giving rise to a somewhat softer EoS at intermediate densities and a stiffer EoS at higher densities than the model without $\Delta$ isobars.

\begin{figure}
\centering 
\includegraphics[width=0.5\textwidth]{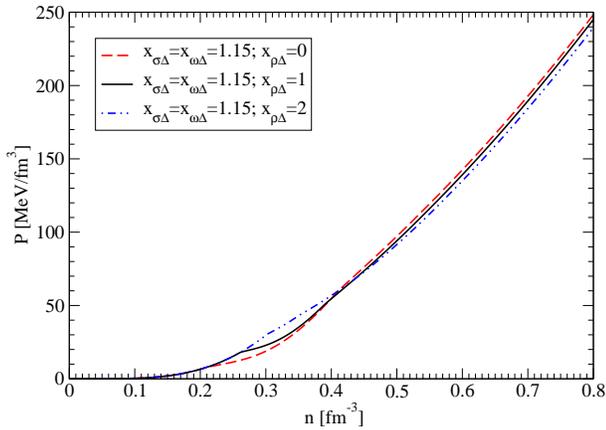}
\caption{Pressure versus baryon density of $\beta$-stable NS matter for the FSU2H model, taking $x_{\sigma \Delta}=x_{\omega \Delta}=1.15$ and different values of the $x_{\rho \Delta}$ coupling.}
\label{fig:eos_rho}
\end{figure}

We now explore the effect of the parameter $x_{\rho \Delta}$, which has been so far kept to $x_{\rho \Delta}=1$. Since this parameter is not constrained by experiments, a few studies \citep{Cai2015CriticalStars,Li:2018qaw} have varied its value within a sizable range. In Fig.~\ref{fig:eos_rho} we show three EoSs that correspond to fixing $x_{\sigma \Delta}=x_{\omega \Delta}=1.15$ and taking $x_{\rho \Delta}=0$, 1, and 2. Since the $\rho$ field is negative, it makes the  $\Delta^-$ less attractive, and hence larger values of $x_{\rho \Delta}$ postpone the appearance of the $\Delta^-$ state to larger densities. We observe that the change of  $x_{\rho \Delta}$ between 0 and 2 affects the EoS mostly in the density range $[0.2-0.4]$ fm$^{-3}$. In this density range, the EoS is softer if $x_{\rho \Delta}$ is lower and, consequently, intermediate size stars acquire a smaller radius as they are easier to compress. At high densities above $\sim$0.4 fm$^{-3}$, the EoS, in contrast, is somewhat stiffer for a lower $x_{\rho \Delta}$ value. The compensating effect between the change of the pressure at densities below and above $\sim$0.4 fm$^{-3}$ leaves the maximum mass of heavy stars almost intact with the modification of $x_{\rho \Delta}$.

\begin{figure}[t]
\centering 
\includegraphics[width=0.5\textwidth]{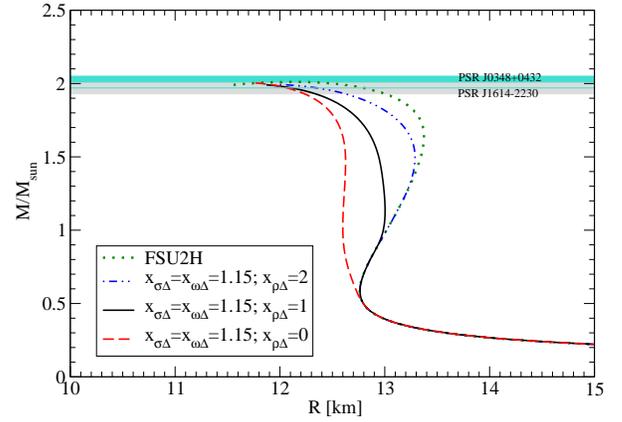}
\caption{Mass versus radius profile for hyperonic NSs in the FSU2H EoS without including $\Delta$ isobars (dotted green line) and for the EoS which includes $\Delta$ degrees of freedom, taking $x_{\sigma \Delta} = x_{\omega\Delta} =1.15$ and  $x_{\rho \Delta} =0$ (dashed red line), $x_{\rho \Delta} =1$ (solid black line), and $x_{\rho \Delta} =2$ (dashed-double dotted blue line). The horizontal shaded regions represent the masses $M = 1.97\pm 0.04 M_{\odot}$ in the pulsar PSR J1614-2230 (gray band) and $M = 2.01\pm 0.04 M_{\odot}$ in the pulsar PSR J0348+0432  (blue band).}
\label{fig:m_r}
\end{figure}

These effects are better seen in Fig.~\ref{fig:m_r}, where we compare the mass-radius relation obtained for the FSU2H EoS (without inclusion of $\Delta$'s) with those obtained taking the strength ratios  $x_{\sigma \Delta} = x_{\omega \Delta} = 1.15$, for which the  2$M_{\odot}$ limit is fulfilled, and $x_{\rho\Delta}=0$, 1 and 2.
We observe that, as the value of $x_{\rho\Delta}$ is reduced, so is the radius of the star, an effect that is magnified for stars with masses around $1.5 M_\odot$ reaching a reduction of 0.7 km, whereas the maximum mass remains at 2$M_{\odot}$ for the three $x_{\rho\Delta}$ values.

\begin{table*}[t]
\centering
\caption{Maximum mass, radius at maximum mass, central density, radius at $M=1.4 M_{\odot}$, and onset densities for the $\Delta^-$ baryon and the $\Lambda$ hyperon. The results are shown for the FSU2H EoS with hyperons but without $\Delta$'s, and for the EoS with $\Delta$'s for the cases $x_{\sigma\Delta} = x_{\omega\Delta} = 0.8$ with $x_{\rho\Delta} =1$, $x_{\sigma\Delta} = x_{\omega\Delta} = x_{\rho\Delta} =1$, and $x_{\sigma\Delta} = x_{\omega\Delta} = 1.15$ with $x_{\rho\Delta} =1,2,0$.}
\label{tab:star_props}
\begin{tabular}{|c|c|c|c|c|c|c|}
\hline
Model & $M_{\text{max}}/M_{\odot}$ & $R(M_{\text{max}})$ (km) & $n_c$ (fm$^{-3}$)& $R(1.4M_{\odot}$) (km) & Onset $n_{\Delta^-}({\rm fm^{-3}})$& Onset $n_{\Lambda} ({\rm fm^{-3}})$\\
\hline 
FSU2H & 2.01 & 12.13 & 0.87 & 13.30 & - & 0.33 \\
\hline 
$x_{\sigma\Delta} = x_{\omega\Delta} = 0.8, x_{\rho\Delta} =1$ & 1.75 & 12.40 & 0.68 & 13.03 & 0.28 & 0.41 \\
\hline 
$x_{\sigma\Delta} = x_{\omega\Delta} = x_{\rho\Delta} =1$ & 1.87 & 12.23 & 0.73 & 13.00 & 0.27 & 0.40 \\
\hline 
$x_{\sigma\Delta} = x_{\omega\Delta} = 1.15, x_{\rho\Delta} =1$ & 2.00 & 11.87 & 0.85 & 12.97 & 0.26 & 0.39 \\
\hline 
$x_{\sigma\Delta} = x_{\omega\Delta} = 1.15, x_{\rho\Delta} =2$ & 2.00 & 12.00 & 0.85 & 13.27 & 0.30 & 0.35 \\
\hline 
$x_{\sigma\Delta} = x_{\omega\Delta} =1.15, x_{\rho\Delta} =0$ & 2.00 & 11.76 & 0.85 & 12.62 & 0.21 & 0.41 \\
\hline 
\end{tabular} 
\end{table*}

In Table~\ref{tab:star_props} we summarize some NS properties obtained with the different models explored in the present paper. In the case of using the same meson couplings as those of the nucleons, 
the presence of $\Delta$ isobars decreases the maximum mass, as well as the radius of a canonical star of $M=1.4M_\odot$, with respect to the EoS without $\Delta$ isobars. Increasing the $x_{\sigma \Delta} $ and $x_{\omega \Delta}$ parameters so as to fulfill the maximum 2$M_{\odot}$ condition, gives rise to more compact stars, especially if a small value for $x_{\rho \Delta}$ is also considered, hence producing canonical NSs with realistic radii \citep{angels}. 

\begin{figure}[t]
\hspace*{-8mm}
\includegraphics[width=0.38\textwidth,angle=-90]{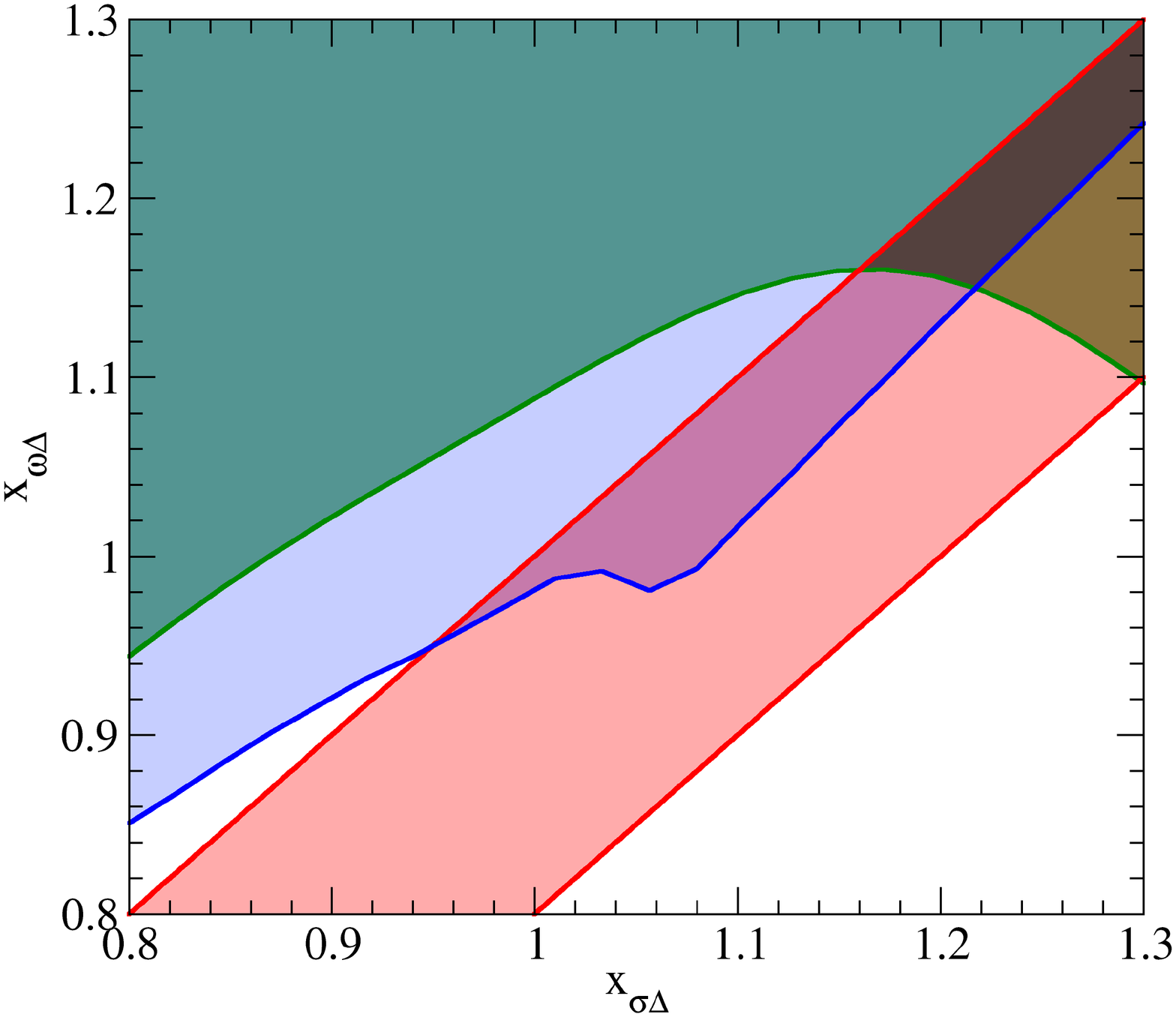}
\caption{Relation between the coupling ratios $x_{\omega \Delta}$ and $x_{\sigma \Delta}$ ($x_{\rho \Delta} = 1$) within the FSU2H  model which lead to stable EoS (shaded blue region) and  to maximum masses higher than $1.998M_{\odot}$ (shaded green region).  The experimental constraints on the difference between $x_{\omega\Delta} $ and $x_{\sigma \Delta}$ \citep{WEHRBERGER1989797}  (shaded red region) are also shown. The dark brown shaded area corresponds to the region where all the constraints are satisfied.}
\label{fig:xsd_xwd}
\end{figure}

In Fig.~\ref{fig:xsd_xwd} we summarize the constraints which the strength ratios $x_{\sigma \Delta}$ and $x_{\omega \Delta}$ must fulfill to ensure the existence of NSs compatible with the stability criterion and with the observational results. Stable EoSs are generated for parameters lying in the shaded blue region. The green region covers the configurations giving maximum NS masses higher than $2M_{\odot}$. The region enclosed between the red lines indicates the combinations of parameters that fulfill the experimental constraints for the difference between $x_{\sigma\Delta} $ and $x_{\omega \Delta}$ \citep{WEHRBERGER1989797}, see Eq.~(\ref{eq:experiment}).
The dark brown region corresponds to the $x_{\sigma \Delta}$ and $x_{\omega \Delta}$ values for which all the constrains are satisfied. We notice that this region implies that the interaction between $\Delta$ isobars and the $\sigma$ and $\omega$ meson fields must be 10\%--30\% larger than in the case of nucleons. For $x_{\sigma\Delta} \simeq x_{\omega \Delta}$ this implies a larger (more negative) $\Delta$ potential than the nucleonic potential in symmetric nuclear matter at normal density. Thus, whereas to date the laboratory experiments are not conclusive on the size of the $\Delta$ potential with respect to the nucleonic one \citep{Ericson:1988gk,Horikawa:1980cv,Nakamura:2009iq,Lehr:1999zr,OConnell:1990njm}, the astrophysical information appears to clearly advocate for a more attractive potential for the $\Delta$ than for the nucleon.

In the next figure, Fig.~\ref{fig:tidal}, we show our results for the mass-weighted tidal deformability
$\tilde{\lambda}$ of a NS binary system [see Eq.~(\ref{tlamb})] as a function of the chirp mass 
$\mathcal{M} = \left(M_1 M_2 \right)^{3/5} / \left(M_1 + M_2 \right)^{1/5}$,
where $M_1$ and $M_2$ are the masses of the two stars of the binary.
The calculations are performed with our baseline EoS, i.e., the FSU2H EoS without $\Delta$ isobars, and with the FSU2H EoS
including the $\Delta$ baryon with couplings $x_{\sigma\Delta}= x_{\omega \Delta}=1.15$ and $x_{\rho\Delta}=0$. 
We plot the predictions for the cases $M_2=0.7 M_1$ and $M_2=M_1$, accounting for the mass asymmetry range compatible with the 
component masses $M_1 \in (1.36,1.60) M_\odot$ and $M_2 \in (1.16,1.36) M_\odot$ reported by the LIGO-Virgo collaboration for the GW170817 
NS merger \citep{Abbott:2018wiz}. For comparison we also display in Fig.~\ref{fig:tidal} the constraint on $\tilde{\lambda}$ from 
the GW170817 event at the measured chirp mass of $\mathcal{M}= 1.186^{+0.001}_{-0.001}M_\odot$ \citep{Abbott:2018wiz}.
The latest analysis of the GW170817 data by the LIGO-Virgo collaboration published in \cite{Abbott:2018wiz},
for the case of low-spin priors (which are more consistent with the spin distribution of the population of galactic NS binaries),
has obtained for this NS merger a dimensionless tidal deformability $\tilde{\Lambda} = 300^{+420}_{-230}$ at 90\% confidence level 
(the new result refines the upper bound $\tilde{\Lambda} \lesssim 800$ reported in the initial data analysis \citep{TheLIGOScientific:2017qsa}). 
Assuming $M_2=0.7M_1$ and the measured $\mathcal{M}$ value, Eq.~(\ref{TLambda}) that relates $\tilde{\Lambda}$ with $\tilde{\lambda}$ 
leads to the constraint on $\tilde{\lambda}$ that we show by the vertical arrow in Fig.~\ref{fig:tidal}
(for $M_2=M_1$ the lower and upper values of this constraint on $\tilde{\lambda}$ are shifted downwards by 10\%).

\begin{figure}[t]
\centering 
\includegraphics[width=0.45\textwidth]{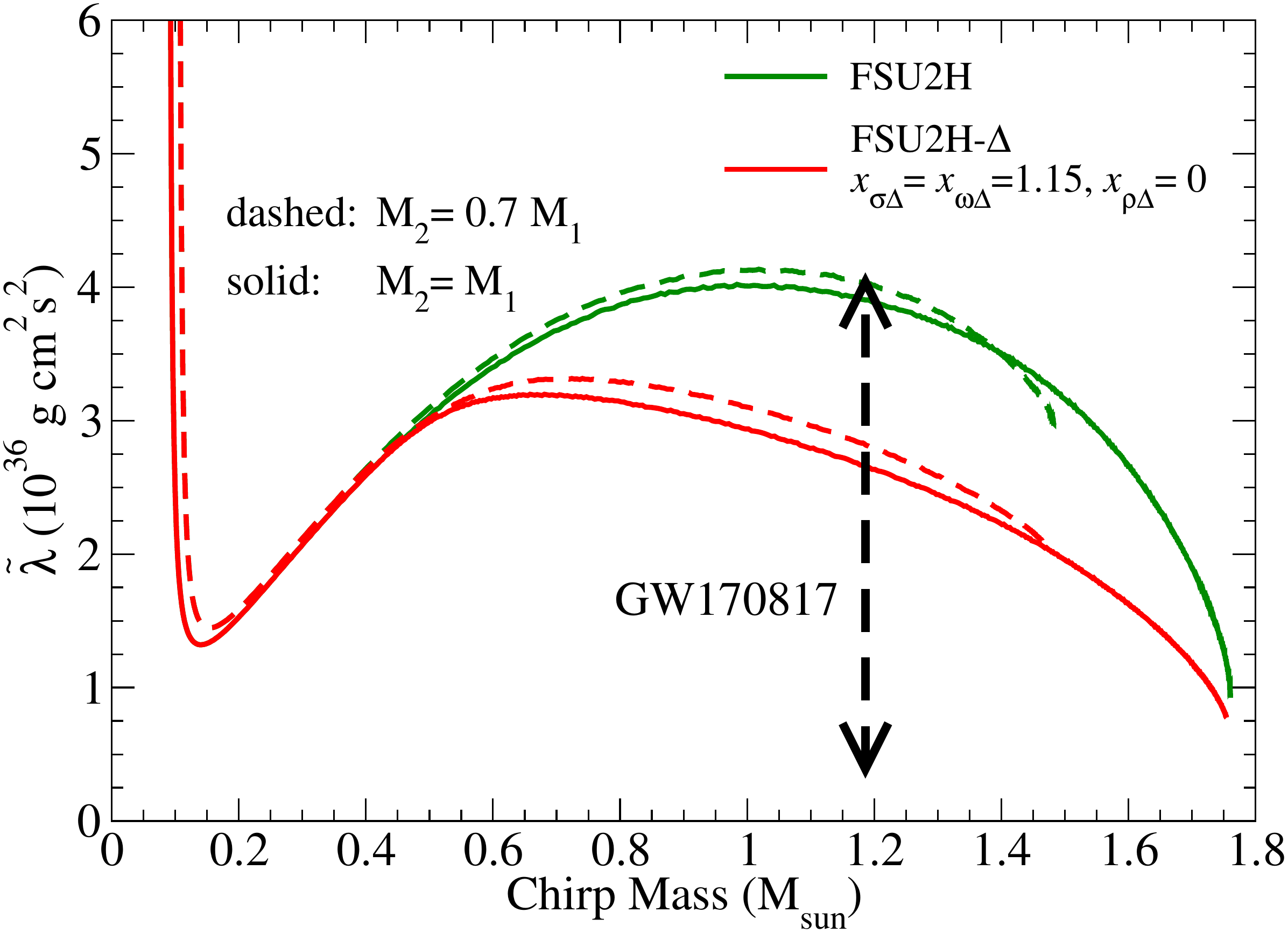}
\caption{Mass-weighted tidal deformability of a NS binary system as a function of the chirp mass, for two different mass ratios of the component stars, calculated for the EoS with hyperons and for the EoS including hyperons and $\Delta$ degrees of freedom with couplings $x_{\sigma \Delta}=x_{\omega \Delta}=1.15$, $x_{\rho \Delta}=0$. The vertical arrow at chirp mass $\mathcal{M}=1.186 M_{\odot}$ shows the constraint estimated from the GW170817 event~\citep{Abbott:2018wiz}.}
\label{fig:tidal}
\end{figure}

We can see in Fig.~\ref{fig:tidal} that the result for $\tilde{\lambda}$ of the EoS
with hyperons but without $\Delta$ isobars lies in the upper boundary of the GW170817 event.
The inclusion of the $\Delta$ baryons has the effect of reducing the mass-weighted tidal deformability
for all values of interest of the chirp mass.
In particular, at the measured chirp mass $\mathcal{M}=1.186 M_{\odot}$ the calculation of the tidal deformability
with the EoS with $\Delta$ isobars is compatible with the GW170817 constraint on $\tilde{\lambda}$.
The improvement is due to the fact that the incorporation of $\Delta$ isobars softens the EoS at $\sim$1.25--2.5 times the saturation density $n_0$. 
This softening is responsible for smaller radii (the Love number $k_2$ is also somewhat smaller
with $\Delta$ isobars) and, hence, it produces smaller tidal deformability values. Our conclusion is in agreement with a
recent study on the presence of $\Delta$ isobars using another type of density functionals that also include
hyperons \citep{Li:2019tjx}. Thus, the GW170817 event is consistent with a merger of a 
binary NS system having hyperons and $\Delta$ isobars in the stellar core.

\begin{figure}[t]
\centering 
\includegraphics[width=0.45\textwidth]{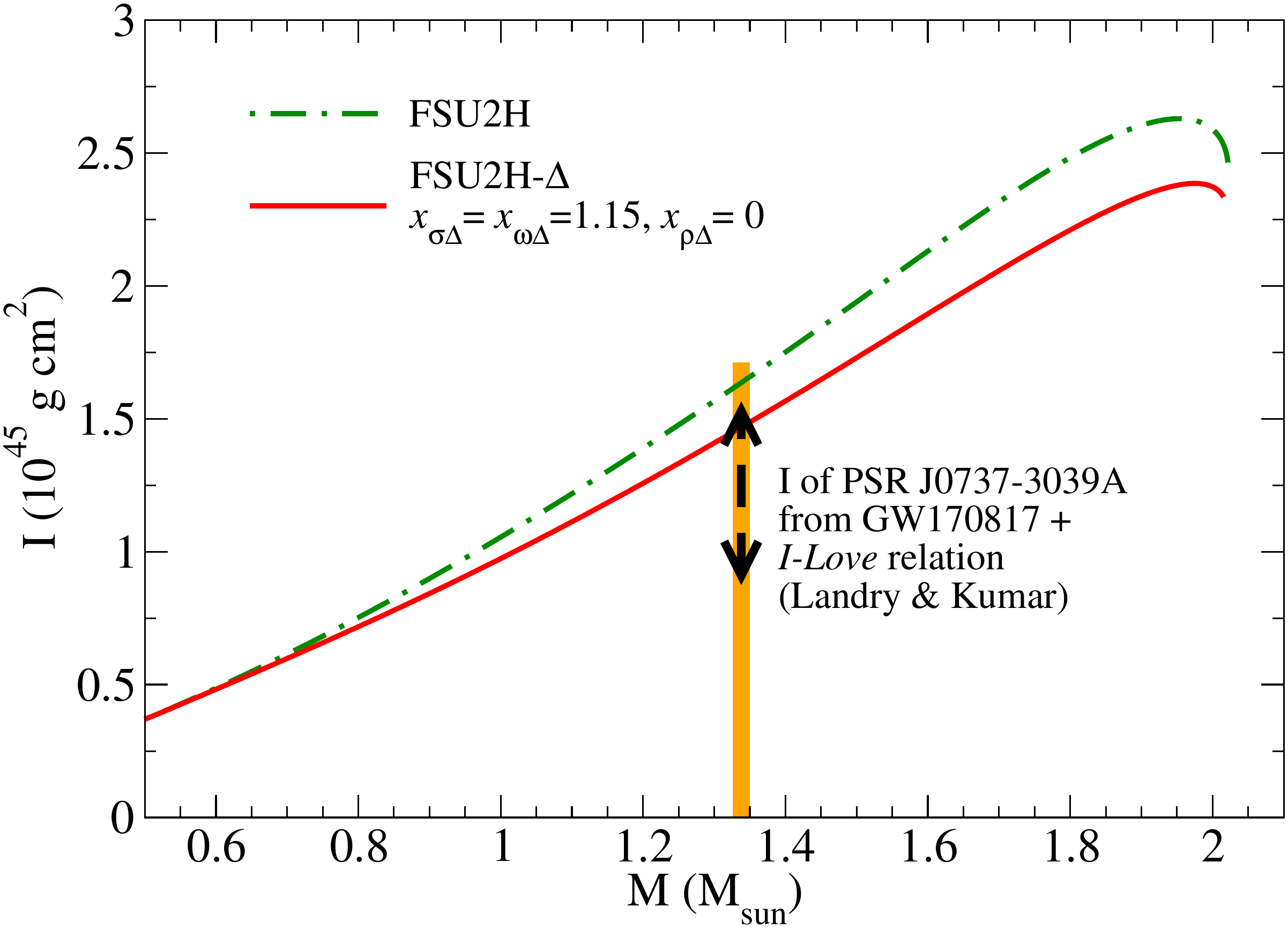}
\caption{Neutron star moment of inertia as a function of the star's mass. The indicated constraints at 1.338$M_{\odot}$ are those of \cite{Landry:2018jyg} for the primary component of the double pulsar system PSR J0737-3039.}
\label{fig:inertia}
\end{figure}

We conclude by displaying in Fig.~\ref{fig:inertia} the NS moment of 
inertia calculated from the EoSs without $\Delta$'s and with $\Delta$'s 
(with couplings $x_{\sigma\Delta}= x_{\omega \Delta}=1.15$ and 
$x_{\rho\Delta}=0$). The results have been obtained by solving Hartle’s 
slow-rotation differential equation for the moment of inertia in general 
relativity coupled to the TOV equations \citep{Hartle1967}. Astronomical 
observations of binary pulsars may provide information on the moment of 
inertia of NSs, ultimately offering possible limits on the underlying EoS 
\citep{Lattimer:2004nj}. The only double-pulsar system known to date is PSR 
J0737-3039. The mass of its primary component PSR J0737-3039A, or pulsar A, 
is of 1.338$M_{\odot}$. It is expected that a precise measurement regarding the 
moment of inertia of this slowly-rotating pulsar will be obtained in the 
near future from radio observations of PSR J0737-3039 \citep{Burgay:2003jj,Lyne:2004cj}.
In a recent work, \cite{Landry:2018jyg} use approximately 
universal relations among NS observables \citep{Yagi:2013bca,Yagi:2013awa}, known as the binary-Love and 
$I$-Love relations (recall that the tidal deformability is related to the 
Love number), to determine a range of $1.15_{-0.24}^{+0.38} \times 10^{45} 
$~g~cm$^2$ for $I$ of pulsar A from the 90\% credible limits on the tidal 
deformability of a 1.4$M_{\odot}$ NS reported by LIGO-Virgo from the 
GW170817 event \citep{Abbott:2018wiz}. \cite{Landry:2018jyg} also facilitate a 
wider range $I \leq 1.67 \times 10^{45} $~g~cm$^2$ for $I$ of pulsar A 
from the less restrictive upper limit on the tidal deformability obtained 
by the LIGO-Virgo collaboration in their initial data analysis of GW170817 in \cite{TheLIGOScientific:2017qsa}.
We have plotted in Fig.~\ref{fig:inertia} these two constraints reported 
in \cite{Landry:2018jyg} for $I$ of pulsar A (we show both of the constraints 
because there may be a certain dependence on the analysis and on the 
assumed boundaries for $\Lambda$ of a 1.4$M_{\odot}$ star).
We can see in Fig.~\ref{fig:inertia} that the moment of inertia is an observable
highly sensitive to the existence of $\Delta$ isobars in the composition of the core of the 
star. The presence of these particles, in particular for smaller values 
of the coupling of the $\Delta$ to the $\rho$-meson, improves the agreement
with the constraints on the moment of inertia of pulsar A, 
whose measurement is anticipated within the next few years.

\section{Conclusions}
\label{sec:conclusions}

We have studied the implications of the appearance of $\Delta$ isobars in the interior of NSs. Within the newly developed RMF model FSU2H \citep{angels,pasa} which incorporates hyperons, we have employed phenomenological analyses on electron-nucleus reactions in the region of the $\Delta$  excitation in order to constrain the couplings between the $\Delta$ isobars and the attractive $\sigma$ meson and the repulsive $\omega$ meson to the range $0 \lesssim x_{\sigma \Delta} - x_{\omega \Delta} \lesssim 0.2$. In this manner, we have analyzed the composition, mass, radius, and tidal deformabilities of NSs as we incorporate the $\Delta$ degrees of freedom. 

The presence of the $\Delta^-$ isobar in the composition of the star systematically postpones the emergence of the $\Lambda$ hyperon to higher densities. Only if one relaxes the constraint on the coupling of the $\Delta$ isobars to the $\sigma$ and $\omega$ mesons, the $\Lambda$ hyperon can appear for lower densities than the $\Delta^-$. In fact, the $\Delta^-$, $\Lambda$ and $\Delta^0$ baryons appear consecutively as the density of the stellar core increases, whereas the $\Delta^+$ and $\Delta^{++}$ baryons enter the composition at even higher densities. This is due to the different $\beta$-stability conditions to be fulfilled by the $\Delta$ charge states, together with the different charge-dependent single-particle potentials. It is to be noted that the $\Delta^-$ partially plays the role of the $\Sigma^-$ and $\Xi^-$ hyperons in compensating the positive charge of the proton, leading to the absence of these hyperons in the composition of the NS in our EoS. Moreover, the simultaneous increase of the coupling ratios $x_{\sigma \Delta}$ and $x_{\omega \Delta}$ gives rise to lower values of the potential for the $\Delta^-$ and the $\Delta^0$ states around densities of 2--$3\, n_0$, allowing for their earlier appearance in the composition, in contrast to a delay of the entrance of the $\Delta^+$ and $\Delta^{++}$ states, as their single-particle potentials become more repulsive for their onset densities above $3\, n_0$.  

In order to fulfill the observations of massive NSs of $2 M_{\odot}$ while having a stable solution for the EoS for all densities studied, we have chosen $x_{\sigma \Delta}=x_{\omega \Delta}=1.15$. Compared to the original EoS without $\Delta$ degrees of freedom, with these values we have obtained a softer EoS at intermediate densities that becomes stiffer at higher densities. As a consequence, NSs with smaller radii are obtained while still reproducing the $2 M_{\odot}$ observations. Indeed, smaller values of the unknown $x_{\rho \Delta}$ coupling between the $\Delta$ isobars and the isovector $\rho$ meson lead to smaller stellar radii compared to those obtained in the original EoS. This is due to the extra softening of the EoS for a smaller $x_{\rho \Delta}$ coupling, as the $\Delta^-$ state becomes more attractive and appears at lower densities. These small values for radii favour smaller tidal deformabilities, more consistent with the value derived from the recent LIGO-Virgo gravitational-wave detection GW170817 accounting for the merger of two NSs \citep{TheLIGOScientific:2017qsa,Abbott:2018exr,Abbott:2018wiz}. We also find that the moment of inertia of slowly-rotating NSs displays a significant dependence on the presence of $\Delta$ isobars in the star's interior, which leads to a reduction of the calculated values for the moment of inertia. 

We have extended our study to determine the parameter space of $x_{\sigma \Delta}$ and $x_{\omega \Delta}$ values compatible with a stable EoS, the $2 M_{\odot}$ observations and the phenomenological analyses on electron-nucleus reactions. We have found that the interaction between the $\Delta$ baryon and the $\sigma$ and $\omega$ fields must be 10--30\% larger than in the case of nucleons, which, in turn, implies that the astrophysical information suggests that the $\Delta$ potential is more attractive than the nucleon potential at normal nuclear matter density.  

We end noting that the triple requirement imposed by large maximum NS masses, small stellar radii and the low value of the tidal deformability derived from the GW170817 binary NS merger is a very demanding challenge for the nuclear models of the EoS, especially when the models include hyperons. In this work we have seen that
the consideration of $\Delta$ degrees of freedom significantly contributes to help reconcile the theoretical EoS with all of the observational constraints.

\vspace{1.cm}

\acknowledgments
L.T. acknowledges support from the FPA2016-81114-P Grant from Ministerio de Ciencia, Innovacion y Universidades, Heisenberg Programme of the Deutsche Forschungsgemeinschaft under the Project Nr. 383452331 and PHAROS COST Action CA16214. A.R., M.C. and C.G. acknowledge support from Grant No. FIS2017-87534-P from MINECO and the project MDM-2014-0369 of ICCUB (Unidad de Excelencia Mar\'{\i}a de Maeztu) from MINECO. C.G. also acknowledges Grant BES-2015-074210 from MINECO. 

\newpage

\bibliography{references.bib}

\begin{thebibliography}{}
\expandafter\ifx\csname natexlab\endcsname\relax\def\natexlab#1{#1}\fi
\providecommand{\url}[1]{\href{#1}{#1}}
\providecommand{\dodoi}[1]{doi:~\href{http://doi.org/#1}{\nolinkurl{#1}}}
\providecommand{\doeprint}[1]{\href{http://ascl.net/#1}{\nolinkurl{http://ascl.net/#1}}}
\providecommand{\doarXiv}[1]{\href{https://arxiv.org/abs/#1}{\nolinkurl{https://arxiv.org/abs/#1}}}

\bibitem[{Abbott {et~al.}(2017)}]{TheLIGOScientific:2017qsa}
Abbott, B.~P., {et~al.} 2017, Phys. Rev. Lett., 119, 161101,
  \dodoi{10.1103/PhysRevLett.119.161101}

\bibitem[{Abbott {et~al.}(2018)}]{Abbott:2018exr}
---. 2018, Phys. Rev. Lett., 121, 161101,
  \dodoi{10.1103/PhysRevLett.121.161101}

\bibitem[{Abbott {et~al.}(2019)}]{Abbott:2018wiz}
---. 2019, Phys. Rev., X9, 011001, \dodoi{10.1103/PhysRevX.9.011001}

\bibitem[{Ahn {et~al.}(2013)}]{Ahn:2013poa}
Ahn, J.~K., {et~al.} 2013, Phys. Rev., C88, 014003,
  \dodoi{10.1103/PhysRevC.88.014003}

\bibitem[{Antoniadis {et~al.}(2013)Antoniadis, Freire, Wex, Tauris, Lynch, van
  Kerkwijk, Kramer, Bassa, Dhillon, Driebe, Hessels, Kaspi, Kondratiev, Langer,
  Marsh, McLaughlin, Pennucci, Ransom, Stairs, van Leeuwen, Verbiest, \&
  Whelan}]{2013Sci...340..448A}
Antoniadis, J., Freire, P.~C.~C., Wex, N., {et~al.} 2013, Science, 340, 448,
  \dodoi{10.1126/science.1233232}

\bibitem[{Banik {et~al.}(2014)Banik, Hempel, \& Bandyopadhyay}]{Banik:2014qja}
Banik, S., Hempel, M., \& Bandyopadhyay, D. 2014, Astrophys. J. Suppl., 214,
  22, \dodoi{10.1088/0067-0049/214/2/22}

\bibitem[{Baym {et~al.}(1971)Baym, Pethick, \& Sutherland}]{Baym:1971pw}
Baym, G., Pethick, C., \& Sutherland, P. 1971, Astrophys. J., 170, 299,
  \dodoi{10.1086/151216}

\bibitem[{Boguta \& Bodmer(1977)}]{Boguta:1977xi}
Boguta, J., \& Bodmer, A.~R. 1977, Nucl. Phys., A292, 413,
  \dodoi{10.1016/0375-9474(77)90626-1}

\bibitem[{Burgay {et~al.}(2003)}]{Burgay:2003jj}
Burgay, M., {et~al.} 2003, Nature, 426, 531, \dodoi{10.1038/nature02124}

\bibitem[{Cai {et~al.}(2015)Cai, Fattoyev, Li, \&
  Newton}]{Cai2015CriticalStars}
Cai, B.~J., Fattoyev, F.~J., Li, B.~A., \& Newton, W.~G. 2015, Physical Review
  C, 92, 1, \dodoi{10.1103/PhysRevC.92.015802}

\bibitem[{Chatterjee \& Vidaña(2016)}]{Chatterjee:2015pua}
Chatterjee, D., \& Vidaña, I. 2016, Eur. Phys. J., A52, 29,
  \dodoi{10.1140/epja/i2016-16029-x}

\bibitem[{Chen \& Piekarewicz(2014)}]{Chen:2014sca}
Chen, W.-C., \& Piekarewicz, J. 2014, Phys. Rev., C90, 044305,
  \dodoi{10.1103/PhysRevC.90.044305}

\bibitem[{Chen \& Piekarewicz(2015)}]{Chen:2014mza}
---. 2015, Phys. Lett., B748, 284, \dodoi{10.1016/j.physletb.2015.07.020}

\bibitem[{Colucci \& Sedrakian(2013)}]{Colucci:2013pya}
Colucci, G., \& Sedrakian, A. 2013, Phys. Rev., C87, 055806,
  \dodoi{10.1103/PhysRevC.87.055806}

\bibitem[{Cromartie {et~al.}(2019)}]{Cromartie:2019kug}
Cromartie, H.~T., {et~al.} 2019.
\newblock \doarXiv{1904.06759}

\bibitem[{Damour \& Nagar(2009)}]{Damour:2009vw}
Damour, T., \& Nagar, A. 2009, Phys. Rev., D80, 084035,
  \dodoi{10.1103/PhysRevD.80.084035}

\bibitem[{Demorest {et~al.}(2010)Demorest, Pennucci, Ransom, Roberts, \&
  Hessels}]{2010Natur.467.1081D}
Demorest, P.~B., Pennucci, T., Ransom, S.~M., Roberts, M.~S.~E., \& Hessels,
  J.~W.~T. 2010, Nature, 467, 1081, \dodoi{10.1038/nature09466}

\bibitem[{Drago {et~al.}(2014{\natexlab{a}})Drago, Lavagno, \&
  Pagliara}]{Drago2014CanExist}
Drago, A., Lavagno, A., \& Pagliara, G. 2014{\natexlab{a}}, Physical Review D,
  89, 1, \dodoi{10.1103/PhysRevD.89.043014}

\bibitem[{Drago {et~al.}(2014{\natexlab{b}})Drago, Lavagno, Pagliara, \&
  Pigato}]{Drago}
Drago, A., Lavagno, A., Pagliara, G., \& Pigato, D. 2014{\natexlab{b}}, Phys.
  Rev., C90, 65809, \dodoi{10.1103/PhysRevC.90.065809}

\bibitem[{Ericson \& Weise(1988)}]{Ericson:1988gk}
Ericson, T. E.~O., \& Weise, W. 1988, {Pions and Nuclei}, Vol.~74 (Oxford, UK:
  Clarendon Press).
\newblock
  \url{http://www-spires.fnal.gov/spires/find/books/www?cl=QC793.5.M42E75::1988}

\bibitem[{Favata(2014)}]{Favata:2013rwa}
Favata, M. 2014, Phys. Rev. Lett., 112, 101101,
  \dodoi{10.1103/PhysRevLett.112.101101}

\bibitem[{Flanagan \& Hinderer(2008)}]{Flanagan:2007ix}
Flanagan, E.~E., \& Hinderer, T. 2008, Phys. Rev., D77, 021502,
  \dodoi{10.1103/PhysRevD.77.021502}

\bibitem[{Fortin {et~al.}(2015)Fortin, Zdunik, Haensel, \&
  Bejger}]{Fortin:2014mya}
Fortin, M., Zdunik, J.~L., Haensel, P., \& Bejger, M. 2015, Astron. Astrophys.,
  576, A68, \dodoi{10.1051/0004-6361/201424800}

\bibitem[{Glendenning(1985)}]{giant}
Glendenning, N.~K. 1985, Astrophys. J., 293, 470, \dodoi{10.1086/163253}

\bibitem[{Glendenning(1997)}]{glen}
---. 1997, {Compact Stars: Nuclear Physics, Particle Physics and General
  Relativity} (Springer)

\bibitem[{{Hartle}(1967)}]{Hartle1967}
{Hartle}, J.~B. 1967, Astrophys. J., 150, 1005, \dodoi{10.1086/149400}

\bibitem[{Hinderer(2008)}]{Hinderer:2007mb}
Hinderer, T. 2008, Astrophys. J., 677, 1216, \dodoi{10.1086/533487}

\bibitem[{Hinderer {et~al.}(2010)Hinderer, Lackey, Lang, \&
  Read}]{Hinderer:2009ca}
Hinderer, T., Lackey, B.~D., Lang, R.~N., \& Read, J.~S. 2010, Phys. Rev., D81,
  123016, \dodoi{10.1103/PhysRevD.81.123016}

\bibitem[{Horikawa {et~al.}(1980)Horikawa, Thies, \& Lenz}]{Horikawa:1980cv}
Horikawa, Y., Thies, M., \& Lenz, F. 1980, Nucl. Phys., A345, 386,
  \dodoi{10.1016/0375-9474(80)90346-2}

\bibitem[{Horowitz \& Piekarewicz(2001)}]{Horowitz:2000xj}
Horowitz, C.~J., \& Piekarewicz, J. 2001, Phys. Rev. Lett., 86, 5647,
  \dodoi{10.1103/PhysRevLett.86.5647}

\bibitem[{Kohno {et~al.}(2006)Kohno, Fujiwara, Watanabe, Ogata, \&
  Kawai}]{Kohno:2006iq}
Kohno, M., Fujiwara, Y., Watanabe, Y., Ogata, K., \& Kawai, M. 2006, Phys.
  Rev., C74, 064613, \dodoi{10.1103/PhysRevC.74.064613}

\bibitem[{Landry \& Kumar(2018)}]{Landry:2018jyg}
Landry, P., \& Kumar, B. 2018, Astrophys. J., 868, L22,
  \dodoi{10.3847/2041-8213/aaee76}

\bibitem[{Lattimer \& Prakash(2016)}]{Lattimer:2015nhk}
Lattimer, J.~M., \& Prakash, M. 2016, Phys. Rept., 621, 127,
  \dodoi{10.1016/j.physrep.2015.12.005}

\bibitem[{Lattimer \& Schutz(2005)}]{Lattimer:2004nj}
Lattimer, J.~M., \& Schutz, B.~F. 2005, Astrophys. J., 629, 979,
  \dodoi{10.1086/431543}

\bibitem[{Lehr {et~al.}(2000)Lehr, Effenberger, \& Mosel}]{Lehr:1999zr}
Lehr, J., Effenberger, M., \& Mosel, U. 2000, Nucl. Phys., A671, 503,
  \dodoi{10.1016/S0375-9474(99)00845-3}

\bibitem[{Li \& Sedrakian(2019)}]{Li:2019tjx}
Li, J.~J., \& Sedrakian, A. 2019, Astrophys. J. Lett., 874, L22,
  \dodoi{10.3847/2041-8213/ab1090}

\bibitem[{Li {et~al.}(2018)Li, Sedrakian, \& Weber}]{Li:2018qaw}
Li, J.~J., Sedrakian, A., \& Weber, F. 2018, Phys. Lett., B783, 234,
  \dodoi{10.1016/j.physletb.2018.06.051}

\bibitem[{Lyne {et~al.}(2004)}]{Lyne:2004cj}
Lyne, A.~G., {et~al.} 2004, Science, 303, 1153, \dodoi{10.1126/science.1094645}

\bibitem[{Miyatsu {et~al.}(2013)Miyatsu, Yamamuro, \&
  Nakazato}]{Miyatsu:2013hea}
Miyatsu, T., Yamamuro, S., \& Nakazato, K. 2013, Astrophys. J., 777, 4,
  \dodoi{10.1088/0004-637X/777/1/4}

\bibitem[{Motta {et~al.}(2019)Motta, Guichon, \& Thomas}]{Motta:2019ywl}
Motta, T.~F., Guichon, P.~A., \& Thomas, A.~W. 2019.
\newblock \doarXiv{1906.05459}

\bibitem[{Mueller \& Serot(1996)}]{Mueller:1996pm}
Mueller, H., \& Serot, B.~D. 1996, Nucl. Phys., A606, 508,
  \dodoi{10.1016/0375-9474(96)00187-X}

\bibitem[{Nakamura {et~al.}(2010)Nakamura, Sato, Lee, Szczerbinska, \&
  Kubodera}]{Nakamura:2009iq}
Nakamura, S.~X., Sato, T., Lee, T. S.~H., Szczerbinska, B., \& Kubodera, K.
  2010, Phys. Rev., C81, 035502, \dodoi{10.1103/PhysRevC.81.035502}

\bibitem[{N{\"a}ttil{\"a} {et~al.}(2016)N{\"a}ttil{\"a}, Steiner, Kajava,
  Suleimanov, \& Poutanen}]{Nattila:2015jra}
N{\"a}ttil{\"a}, J., Steiner, A.~W., Kajava, J. J.~E., Suleimanov, V.~F., \&
  Poutanen, J. 2016, Astron. Astrophys., 591, A25,
  \dodoi{10.1051/0004-6361/201527416}

\bibitem[{Negreiros {et~al.}(2018)Negreiros, Tolos, Centelles, Ramos, \&
  Dexheimer}]{Negreiros:2018cho}
Negreiros, R., Tolos, L., Centelles, M., Ramos, A., \& Dexheimer, V. 2018,
  Astrophys. J., 863, 104, \dodoi{10.3847/1538-4357/aad049}

\bibitem[{O'Connell \& Sealock(1990)}]{OConnell:1990njm}
O'Connell, J.~S., \& Sealock, R.~M. 1990, Phys. Rev., C42, 2290,
  \dodoi{10.1103/PhysRevC.42.2290}

\bibitem[{Ozel {et~al.}(2016)Ozel, Psaltis, Guver, Baym, Heinke, \&
  Guillot}]{Ozel:2015fia}
Ozel, F., Psaltis, D., Guver, T., {et~al.} 2016, Astrophys. J., 820, 28,
  \dodoi{10.3847/0004-637X/820/1/28}

\bibitem[{{\"Ozel, Feryal and Freire, Paulo}(2016)}]{Ozel:2016oaf}
{\"Ozel, Feryal and Freire, Paulo}. 2016, Ann. Rev. Astron. Astrophys., 54,
  401, \dodoi{10.1146/annurev-astro-081915-023322}

\bibitem[{P.~Haensel A.Y.~Potekhin(2007)}]{haensel}
P.~Haensel A.Y.~Potekhin, D. G.~Y. 2007, {Neutron Stars 1. Equation of State
  and Structure}, Vol. 326 (Springer)

\bibitem[{Postnikov {et~al.}(2010)Postnikov, Prakash, \&
  Lattimer}]{Postnikov:2010yn}
Postnikov, S., Prakash, M., \& Lattimer, J.~M. 2010, Phys. Rev., D82, 024016,
  \dodoi{10.1103/PhysRevD.82.024016}

\bibitem[{Provid\^{e}ncia {et~al.}(2019)Provid\^{e}ncia, Fortin, Pais, \&
  Rabhi}]{Providencia:2018ywl}
Provid\^{e}ncia, C., Fortin, M., Pais, H., \& Rabhi, A. 2019, Front. Astron.
  Space Sci., 6, 13, \dodoi{10.3389/fspas.2019.00013}

\bibitem[{Schaffner \& Mishustin(1996)}]{Schaffner:1995th}
Schaffner, J., \& Mishustin, I.~N. 1996, Phys. Rev., C53, 1416,
  \dodoi{10.1103/PhysRevC.53.1416}

\bibitem[{Serot \& Walecka(1986)}]{Serot:1984ey}
Serot, B.~D., \& Walecka, J.~D. 1986, Adv. Nucl. Phys., 16, 1

\bibitem[{Shapiro \& Teukolsky(1983)}]{Shapiro:1983du}
Shapiro, S.~L., \& Teukolsky, S.~A. 1983, {Black holes, white dwarfs, and
  neutron stars: The physics of compact objects}

\bibitem[{Tolos {et~al.}(2017{\natexlab{a}})Tolos, Centelles, \&
  Ramos}]{angels}
Tolos, L., Centelles, M., \& Ramos, A. 2017{\natexlab{a}}, Astrophys. J., 834,
  3, \dodoi{10.3847/1538-4357/834/1/3}

\bibitem[{Tolos {et~al.}(2017{\natexlab{b}})Tolos, Centelles, \& Ramos}]{pasa}
---. 2017{\natexlab{b}}, Publ. Astron. Soc. Austral., 34, e065,
  \dodoi{10.1017/pasa.2017.60}

\bibitem[{Wehrberger {et~al.}(1989)Wehrberger, Bedau, \&
  Beck}]{WEHRBERGER1989797}
Wehrberger, K., Bedau, C., \& Beck, F. 1989, Nuclear Physics A, 504, 797,
  \dodoi{http://dx.doi.org/10.1016/0375-9474(89)90008-0}

\bibitem[{Weissenborn {et~al.}(2012)Weissenborn, Chatterjee, \&
  Schaffner-Bielich}]{Weissenborn:2011ut}
Weissenborn, S., Chatterjee, D., \& Schaffner-Bielich, J. 2012, Phys. Rev.,
  C85, 065802, \dodoi{10.1103/PhysRevC.85.065802, 10.1103/PhysRevC.90.019904}

\bibitem[{Yagi \& Yunes(2013{\natexlab{a}})}]{Yagi:2013bca}
Yagi, K., \& Yunes, N. 2013{\natexlab{a}}, Science, 341, 365,
  \dodoi{10.1126/science.1236462}

\bibitem[{Yagi \& Yunes(2013{\natexlab{b}})}]{Yagi:2013awa}
---. 2013{\natexlab{b}}, Phys. Rev., D88, 023009,
  \dodoi{10.1103/PhysRevD.88.023009}

\bibitem[{Zhu {et~al.}(2016)Zhu, Li, Hu, \& Sagawa}]{Zhu:2016mtc}
Zhu, Z.-Y., Li, A., Hu, J.-N., \& Sagawa, H. 2016, Phys. Rev., C94, 045803,
  \dodoi{10.1103/PhysRevC.94.045803}

\end{thebibliography}
\bibliographystyle{aasjournal}

\end{document}